\documentstyle{elsart}

\begin{document}

\renewcommand{\baselinestretch}{1.5}

\small{\noindent PASPS 2000 invited talk, \today.}
\vspace{1cm}

\begin{frontmatter}

\title{Theory of  Coherent Optical Control of Exciton Spin Dynamics in
a Semiconductor Dot}

\author{Pochung Chen, C. Piermarocchi, and L.J. Sham\thanksref{1}}
\address{Department of Physics, University of California San Diego,
\\   La Jolla, CA 92093-0319, USA}
\thanks[1]{Corresponding author: L.J. Sham, Department of Physics,
University of California San Diego, 9500 Gilman Drive, La Jolla, CA
92093-0319, USA.  Email: lsham@ucsd.edu, Phone: 858 534-3269, Fax: 858
534-2232.}

\date{\today}

\begin{abstract}

We use the spin-polarized  excitons in a single quantum dot to design
optical controls for basic operations in quantum computing.  We
examine the ultrafast nonlinear optical processes required and use
the  coherent nonlinear optical responses to deduce if such processes
are physically reasonable.  The importance and construction of an
entangled state of polarized exciton states in a single quantum dot
is explained.  We put our proposal in perspective with respect to a
number of theoretical suggestions of utilizing the semiconductor
quantum dots.

\vspace{2.5mm}
\end{abstract}

\end{frontmatter}

\newpage
\section{Introduction}

Recently there has been remarkable experimental progress in optical
control of exciton spins in semiconductor heterostructure
\cite{awsch}.  In particular, optical experiments on a single quantum
dot have been demonstrated
\cite{gammon,steel,chavez}.  It is thus an opportune time to consider
in theory what optical processes on the exciton spins in one or a few
quantum dots can lead to potential device applications.  To focus the
discussion, we choose to use quantum computing \cite{preskill,ekert}
as a paradigm.  This approach has the benefit of not only raising
relevant basic issues of optical control of spins for quantum
computing but it may also be relevant to more semiclassical devices.
There are possible extensions of these fundamental quantum processes
to more semiclassical results for an ensemble of dots.  Examples
 include the extension to quantum dots of spin oscillation
effects  \cite{awsch2} or spin-dependent conditional processes
\cite{tack} for an ensemble of excitons in a quantum well which
depend on exciton-exciton interaction \cite{oss}.

In this paper, we start with the simplest example, a model of two
interacting excitons of opposite spins in a single quantum dot and
investigate what is an appropriate quantum bit of information (qubit)
and how, in principle, to address the bit, to form entangle states
and to make a logic gate of two qubits, all with ultrafast
coherent optical control.  These are some of the basic optoelectronic
issues involved in making a device for quantum computing.  We shall
also show how a coherent nonlinear spectrum may help in reaching
these goals.  We have chosen the simplest possible system to increase
the likelihood of an experimental demonstration.

We need to put our work in perspective in relation with the large
number of suggestions for the implementation of quantum computing in
the semiconductor media.  The earliest suggestions use either two
electrons in two connected dots \cite{barenco} or two electron spins
in two dots with exchange coupling
\cite{ld}, controlled respectively by infrared light \cite{sanders}
or by magnetic field.  Another class makes use of the nuclear spins
or electron spins of the implanted impurities to make a transistor
\cite{kane,ucla}.  Yet another suggests the interaction of dots via
the electromagnetic field in a cavity \cite{ima,wang}.  There is one
on optical control of excitons transferring between neighboring dots
\cite{reina}.  The group of proposals closest to ours uses optical
control of interacting excitons in quantum dots \cite{moli,rossi}.
There are interesting differences among the three which will be
discussed below.  They also have in common with the approach of
optically driven evolution of a multilevel atom \cite{rama}, although
the difficulties with atoms as briefly discussed in
Ref.~\cite{barenco} are absent in the dots.

\section{Basic requirements of quantum computing}

First, we need two quantum states to represent the 0 and 1 of a qubit
\cite{preskill}.  Examination of what operations on $n$ pairs of
states are needed to carry out a specific quantum computation, such
as the solution of the Deutsch problem \cite{deutsch}, factorization
\cite{ekert} or a quantum search \cite{grover}, then gives us a target
to design the nonlinear optical operations on the quantum dots.  In
general, the operations fall into three groups of steps
(see Figure~\ref{circuit}).  The first group prepares the system of
$n$ qubits into an entangled (often maximally entangled) state.  This
can be carried out with  separate operations on each qubit.  The
reason for the entanglement is that the next group of operations is
effectively a transformation into another state.  This group of
operation is named the Oracle.  For those of us familiar with
Fortran, CALL SUBROUTINE may be a more transparent analogy. The
advantage of the quantum computing over the classical counterpart
lies in this transformation being sort of a parallel processor for
different possibilities.  If your input state to the Oracle involves
a small portion of the qubits, you are leaving out a lot of
possibilities and thus, are not likely to get your money's worth in
the answer by the Oracle.  Basic to this second group of operations is
the conditional dynamics.  It has been shown that combinations of the
control-not (C-NOT) gate between two qubits and single bit operations
are sufficient to carry out any quantum algorithm \cite{ekert}.  We
shall explain the C-NOT gate in terms of the quantum dot.  The third
group of operations disentangles the output state, which can then be
measured.

Nuclear magnetic resonance experiments have been performed to
implement the solution of the Deutsch problem in its simplest form
\cite{nmr1,nmr2}.  It is instructive to compare the procedure in the
quantum dot with the NMR experiments.

\section{Antiparallel-spin Excitons as Qubits}
\label{biex}

Consider first a simple model of a quantum dot which is a pill box
with the shortest dimension along the growth axis (the $z$ axis).  It
houses the ground state $|0\rangle$, the spin-polarized heavy-hole
exciton state $|+\rangle$ which can be excited by an electromagnetic
wave of circular polarization
$\sigma+$, the  heavy-hole exciton state $|-\rangle$ associated with
the wave of circular polarization $\sigma-$, and the antiparallel spin
biexciton state $|+-\rangle$, as shown in the inset of
Figure~\ref{nls}.  Both
single exciton states have energy $E_X$ and the biexciton energy is
$2E_X - \Delta$, where $\Delta$ is the biexciton binding energy.

We use the absence and presence of each single exciton as a qubit.
Thus $|00\rangle$ which denotes the 0 state for both qubits
corresponds to the ground state $|0\rangle$ and $|10\rangle$,
$|01\rangle$ and $|11\rangle$ correspond to the physical states
$|+\rangle$, $|-\rangle$ and $|+-\rangle$.

A single qubit operation such as flipping the first bit from 0 to 1
requires a sum of two optical excitations $|0\rangle
\rightarrow |+\rangle$,
$|-\rangle \rightarrow |+-\rangle$. This corresponds to the matrix
transformation
\begin{equation} X_1 =\left[\begin{array}{cccc} 0 & 1 & 0 & 0 \\ 1 &
0 & 0 & 0 \\ 0 & 0 & 0 & 1 \\0 & 0 & 1 & 0
\end{array}\right].
\end{equation}
To accomplish this, we might use two sharp resonance
excitations with frequencies corresponding to $E_X$ and
$E_X-\Delta$.  Conceptually, this requires two Rabi flops
\cite{cohen} or, to borrow the spin language \cite{slichter}, a
$\pi-$pulse of each frequency.

A C-NOT operation changes, say, the second qubit (the target)
depending on the state of the first qubit (the control).  If the
first is 0, the second is unchanged and if the first is 1, the second
bit is flipped. The matrix transformation is
\begin{equation} X_1 =\left[\begin{array}{cccc} 1 & 0 & 0 & 0 \\ 0 &
0 & 0 & 1 \\ 0 & 0 & 1 & 0 \\0 & 1 & 0 & 0
\end{array}\right].
\end{equation}
This is accomplished by a single $\pi-$pulse, resulting in $|+\rangle
\leftrightarrow |+-\rangle$.

The ability to address resonantly only the transition between
the exciton and biexciton states without affecting the transition
between the ground state and the single exciton state is rooted in the
strong interaction between two excitons.  In the limit of no
interaction between two excitons, a resonant excitation or rotation of
one pair of states would give rise to an equal amount on the other
pair \cite{cundiff}.  There would then be no possibility of
conditional exciton spin dynamics.

The proposal of resonance addressing of two levels for a C-NOT
operation in a two-qubit system with sufficiently strong interaction
was first put forward by Barenco et al.\ \cite{barenco} for
intrasubband transitions for electrons in the conduction band of two
coupled dots.  Such operations in the optical range is, in the
current state of the art, simpler than in the infrared range
needed for the intrasubband transitions.  The working principle for
two strongly interacting qubits is different from that for two weakly
interacting ones as is used in NMR \cite{nmr1,nmr2}.  In the latter,
the target spin is allowed to precess under the spin-spin interaction
due to the control spin.  The time it takes is much longer than the
Rabi oscillation time scale for addressing each spin individually.
The NMR system can afford such long times since they are still
shorter than the dephasing time of the spins in the millisecond
range.  By contrast, the optical dephasing time is about 40~ps
\cite{steel}.  This can only be accommodated by a judicious design of
a sequence of femtosecond pulses.

The fundamental experimental requirement for being able to carry out
one and two qubit operations is to demonstrate Rabi flops in a
quantum dot.  Rabi oscillation has long been established for the spin
\cite{slichter} and the atomic system \cite{cohen} but has not been
established experimentally in a quantum dot.  In the quantum well,
Rabi oscillations are claimed to have been seen \cite{schulzgen}.
However, since it is possible to excite a large number of extended
excitons, a likely scenario is the oscillation of the coherent
exciton density or the average polarization, as indicated
by the theoretical analysis of the experiment in the same paper.  It
is driven by the electric field, unlike the collective density
oscillation of the exciton gas. \cite{ostreich}.  Thus, it might be
a coherent mixture of the Rabi rotations between pairs of
multiexcitonic states with roughly equal transition frequencies.
If so, this finding in a quantum well could be a harbinger of the
Rabi oscillations in a quantum dot when the excitonic states become
discrete.

\section{Coherent Nonlinear Spectra}

Figure~\ref{nls} shows the coherent nonlinear spectra for two cw laser
beams, with a strong pump beam with $\sigma-$ circular polarization
and with a weak probe beam with $\sigma+$ polarization. The top curve
is the  linear absorption spectrum of the $\sigma+$ probe when the
pump is turned off.  The others are the corresponding spectra when the
$\sigma-$ pump beam is focused at the exciton resonance and two
frequencies below resonance as indicated in energy units on the right
hand margin.  The damping constant is taken to be 15~$\mu$eV,
approximately that of the measured 20~$\mu$eV \cite{steel}.  The pump
beam intensity corresponds to a Rabi energy of 0.05~meV.  It is taken
to be much stronger than was used in the nonlinear measurement
\cite{steel} in order to produce a discernable Rabi splitting as
shown in the lower three curves where the resonance exciton line in
the linear absorption spectrum is now split into two lines about
twice the Rabi energy apart.  The additional doublet 1~meV below the
exciton line is the Rabi splitting of the exciton to biexciton
transition, $|-\rangle \rightarrow |+-\rangle$.  Our use of the 1~meV
biexciton binding energy is conservative compared with the 4~meV
found for the dot used by Steel's group \cite{steel2}.

The coherent cross-polarized pump and probe spectroscopy in a dot
represents the mesoscopic analogous of the Autler-Townes spectroscopy
in atomic systems. Therefore, the presence of the four peaks in the
nonlinear response can be interpreted in a dressed exciton picture,
similar to the well known dressed-atom picture \cite{cohen}.  The
strong $\sigma_-$ pump produces the Rabi oscillation between the two
states $|0\rangle\otimes|N\sigma_-\rangle$ and
$|-\rangle\otimes|(N-1)\sigma_-\rangle$, where $|N
\sigma_-\rangle$ represent the state of the radiation field
with $N$ photons with $\sigma_-$ polarization. Due to this
strong coupling two dressed states result that can be written in the
form:
\begin{eqnarray}
|lower\rangle=\cos\theta|0\rangle\otimes|N
\sigma_-\rangle-\sin\theta|-\rangle\otimes|(N-1)\sigma_-\rangle
\\
|upper\rangle=\sin\theta|0\rangle\otimes|N\sigma_-\rangle
+ \cos\theta|-\rangle\otimes|(N-1)\sigma_-\rangle~,
\end{eqnarray}
with the angle $\theta$ defined as
\begin{equation}
 \tan(2\theta)=-\frac{\sqrt{(\Omega)^2+(\delta)^2}}{\delta} \; ,
\end{equation}
$\delta =\omega_{\mbox{pump}}-E_X/\hbar$ being the detuning
frequency.
  The energy of the lower and upper states are split by $
2\hbar\sqrt{(\Omega)^2+(\delta)^2}$, where $\hbar\Omega$ is the Rabi
energy of the pump field.

The two peaks at the excitonic energy in one of the nonlinear spectra
are due to transitions from the $|lower\rangle\rightarrow |+,N
\sigma_-\rangle$ and $|upper\rangle \rightarrow |+,N \sigma_-\rangle$
by absorption of a $\sigma_+$ photon from the probe. The two peaks at
the minus biexciton binding energy originate from
transitions $|lower\rangle \rightarrow |+-,(N-1) \sigma_-\rangle$ and
$|upper\rangle \rightarrow |+-,(N-1)\sigma_-\rangle$.
The characteristic shape of the doublet in presence of detuning can
also be interpreted within the dressed exciton picture. At resonance
pumping, the Rabi oscillation gives a fifty-fifty probability of
having the system in the lower or upper states, giving equal intensity
to the two peaks of the Rabi splitting. On the other hand, when the
pump has negative detuning,  the full inversion of the $|0\rangle$ and
$|-\rangle$ levels is not realized, and the system will more likely
be in the lower state, which is the one with a bigger $|0\rangle$
component. This gives a bigger intensity for the peak corresponding
to the $|lower\rangle\rightarrow |+,N
\sigma_-\rangle$ transition (the higher energy peak in the doublet).
In presence of a positive detuning the situation is reversed: the
state $|lower\rangle$ is the one at higher energy and the lower energy
peak in the Rabi doublet dominates. (This case is not shown in the
figure.)

The coherent nonlinear spectra are useful in designing the laser
pulses for the time dependent operations to carry out the processes
whose basic units are described in the last section.  Because of
space limitation, we shall publish the time sequence work in
another paper \cite{chen}.  The spectra give the frequencies for the
resonance operations.  The Rabi splitting would give an indication of
sufficient electric field strength for Rabi oscillations.  The pulse
width and biexciton binding energy would provide the limitation of
the pulse width to be used in the qubit operations within the
confines of the dephasing time and of the error caused by affecting
levels not intended to be in the operation.

\section{An Entangled Exciton State}

We have explained how an entangled exciton state is an important
asset in quantum computing.  It is also an asset in quantum
information in general, especially in cryptography and teleportation
\cite{preskill}.  In the scheme for general entanglement in the last
section, we require the Rabi oscillation.  The collaboration of the
Steel group and the Gammon group has experimentally achieved the
entanglement between the two single exciton states $|+\rangle$ and
$|-\rangle$ \cite{gang} without Rabi oscillations.  In the pump and
probe spectroscopy described in the last section, the two laser
beams of opposite circular polarizations are
phase-locked and kept at sufficiently low intensity to be in the
third order regime.  A single exciton state produced by one of the
beams can be coupled to the single exciton state of opposite
polarization by the second order process of de-excitation by the same
beam to the ground state and excitation by the other beam to the
second exciton state. It was found experimentally that the dephasing
is due mostly to recombination with negligible contribution from pure
dephasing mechanisms.  This leads to a very stable entangled state
within the 50~ps time scale. This use of the spin coherence between
two excitons (also called the Zeeman coherence
\cite{hanle}) is an interesting contrast to the optical coherence
used in the Rabi oscillations.  Both types of coherence exist in
spin-polarized excitons \cite{jmmm}.

The resultant state
\begin{equation}
|\Psi\rangle = C_0 |0\rangle + C_+ |+\rangle + C_- |-\rangle
\end{equation}
still involves the ground state with the coefficients $|C_{\pm}|$ of
the order 0.3.  It contains negligible amount of the biexciton state
because of the resonance condition on the exciton energy.   This is a
strong interaction effect in the sense explained in Sec.~\ref{biex}.
The linear combination of the two exciton states is an entanglement
for one of two reasons below.  In our two-qubit system, this is the
combination $|10\rangle$ and $|01\rangle$.  Alternatively, if the
exciton is regarded as a pair of conduction electron and valence
hole, the linear combination of two excitons of opposite polarization
is an EPR state for the electron and hole, each capable of one of two
spin states.

The existence of the entangled state is deduced from the measured
coherent pump and probe spectra.  A Tesla field is used to split the
two spin exciton states.  As the pump beam of one circular
polarization is detuned below the exciton resonance of that
polarization, the probe spectra of the exciton of opposite
polarization acquired a change in line shape, indicating an
interference effect between the third order transition involving the
same exciton state and the third order process described above from
the ground state to the exciton of the pump polarization and then to the
exciton of opposite polarization via the ground state.  The
interaction between the two excitons which form the biexciton states
prevents a similar third order process via the biexciton state to
cancel the one via the ground state.  The cancellation would be
complete in the absence of interaction \cite{gang,cundiff} and the
pump would have no effect on the probe beam.

\section{Discussions}

We have given a theoretical prescription of the optical processes to
build the basic operations for quantum computing in a two qubit system
using two excitons of opposite spins in a single quantum dot.
The advantage lies in the simplicity of the theory which could lead to
a certain degree of simplicity for the experiment.

The proposal is very similar to that of Troiani et al. \cite{moli}
and of  Biolatti et al. \cite{rossi} in the use of optical control in
quantum dots.  It differs from both in using excitons of opposite
spins.  At the two qubit level, our proposal is simpler since any
interference between light of opposite polarization must come from
the exciton-exciton interaction.  Undesired interference effects
between two color light beams of the same polarization via
noninteracting sources may need to be ameliorated.  Nonetheless, this
is an important problem because we would encounter it, if we wish to
increase the number of qubits in a single dot.

This brings us to the question of scaling.  Clearly, while we could
increase the number of qubits in a single quantum dot, the process
would terminate rather quickly.  We could consider two avenues to
scale our two qubit systems to many dots, either by interdot
interaction or by using the cavity as a data bus \cite{wang,ima}.
The exciton interaction can be strong enough for the biexciton to
be seen directly in the interaction with the cavity mode \cite{biex}.
The required interaction between dots needs to be as strong as the
intradot interaction (a few meV's). Thus a scheme of enhancing the
interdot interaction, such as that of Biolatti et al.
\cite{rossi}, might be followed.

\section*{Acknowledgments}

LJS thanks Drs. D. Steel, Gang Chen and D. Gammon for stimulating
discussions. CP acknowledges the support by the Swiss National
Foundation for Scientific Research. This work was supported in
part by the NSF Grant No. DMR 9721444 and in part by DARPA/ONR
N0014-99-1-109.

\newpage

\section*{}

\begin{figure}
\section*{Figure Captions}
\caption{Circuit diagram for a generic quantum computation.  The
three qubit number controls via a function in the subroutine the
change in the target qubit.  The squares denote single qubit
operations.  The Oracle involves at least one C-NOT operation.
\label{circuit}}
\end{figure}

\begin{figure}
\caption{Coherent nonlinear spectra.  The vertical axis is the
intensity of absorbed light in arbitrary units.  The horizontal axis
is $E_{probe}$, the frequency of the probe beam times $\hbar$ measured
from the exciton energy.  The damping constant is $\Gamma = 15
~\mu$eV.  The biexciton binding energy is 1~meV.  The inset depicts
the energy levels in a dot.
\label{nls} }
\end{figure}

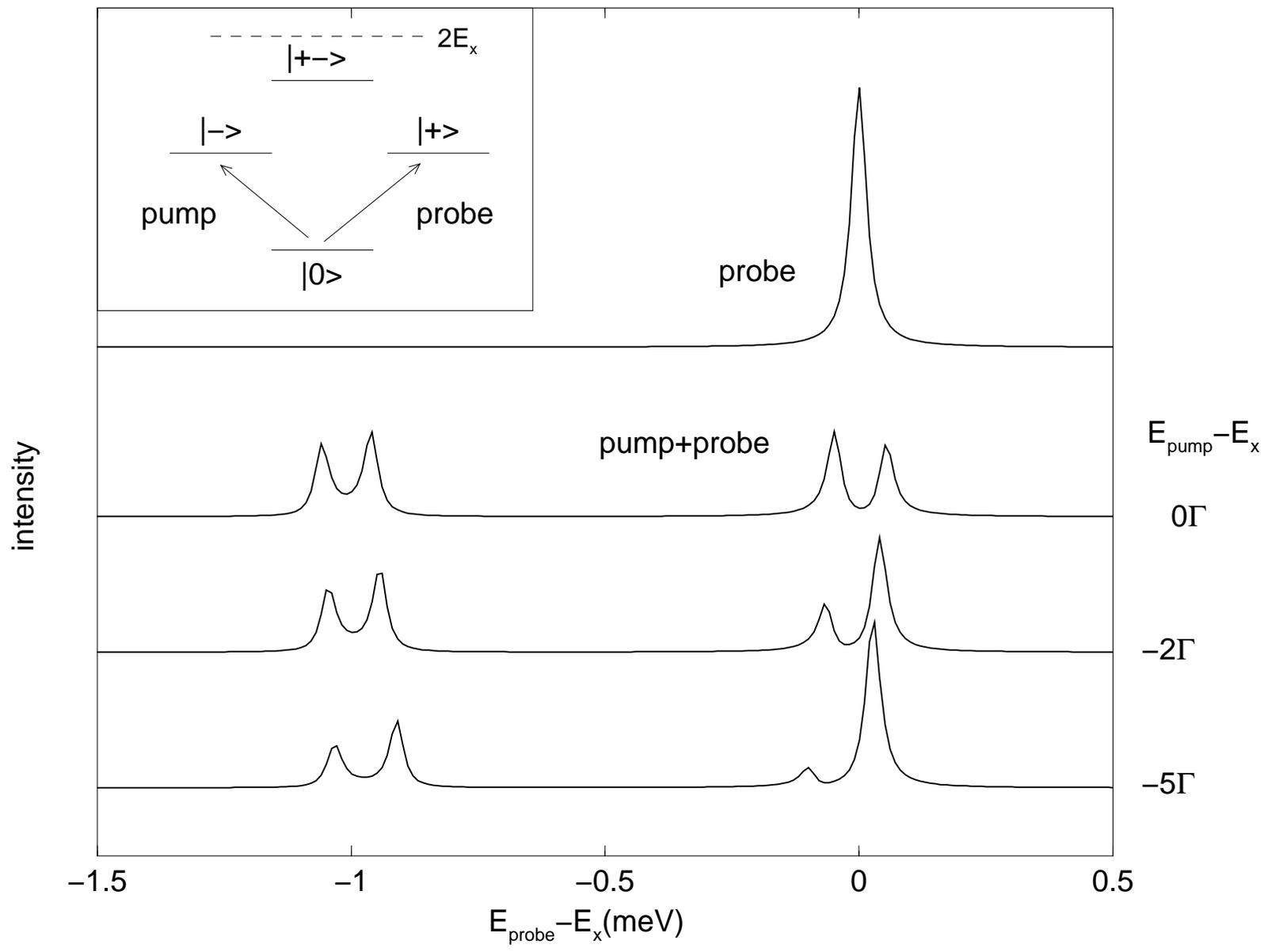
\end{document}